\documentclass[conference]{IEEEtran}
\IEEEoverridecommandlockouts

\usepackage{cite}
\usepackage{amsmath,amssymb,amsfonts}
\usepackage{algorithmic}
\usepackage{graphicx}
\usepackage{textcomp}
\usepackage{xcolor}
\usepackage{subcaption}
\usepackage{multirow}
\usepackage{mathtools}
\usepackage{booktabs}
\usepackage{placeins}
\usepackage{adjustbox}
\usepackage[most]{tcolorbox}
\usepackage{float}
\usepackage{url}

\newcommand{\Design}[0]{\textsc{COPA}}

\tcbset{
  copaexample/.style={
    colback=gray!5,
    colframe=black,
    size=fbox,
    left=1mm,
    right=1mm,
    top=1mm,
    bottom=1mm,
    boxsep=1mm,
    sharp corners,
    breakable,
    width=\linewidth,
    fontupper=\small
  }
}

\def\BibTeX{{\rm B\kern-.05em{\sc i\kern-.025em b}\kern-.08em
    T\kern-.1667em\lower.7ex\hbox{E}\kern-.125emX}}

\begin{document}

\title{COPA: Continual Preference Optimization for Adaptive Prompt Injection Defense}


\author{
    \IEEEauthorblockN{
        Roshan Sood\textsuperscript{1},
        Onat Gungor\textsuperscript{2},
        Tajana Rosing\textsuperscript{1}
    }
    \IEEEauthorblockA{
        \textsuperscript{1}University of California, San Diego, CA, USA \\
        \textsuperscript{2}West Virginia University, Morgantown, WV, USA \\
        \{rosood, tajana\}@ucsd.edu, onat.gungor@wvu.edu
    }
}


\maketitle

\begin{abstract}
LLMs remain vulnerable to prompt injection attacks, where adversarial instructions embedded in user inputs or external content manipulate model behavior and bypass safeguards. Existing defenses are predominantly static, relying on fixed alignment objectives or attack-specific filtering mechanisms that require redesign as new attack strategies emerge. While recent lifelong alignment methods address shifting user preferences, they do not account for adaptive adversaries that continually evolve to exploit weaknesses in previously learned defenses. This limitation is particularly important in real-world deployments, where evolving attack distributions necessitate continual adaptation without sacrificing robustness to previously encountered threats. We present \Design{}, a continual preference optimization framework that treats prompt-injection defense as a lifelong learning problem. Instead of one-time alignment, \Design{} incrementally incorporates feedback from newly observed attacks via GRPO-based optimization and uses margin-weighted experience replay to retain defenses against prior attack classes. This enables continuous adaptation to emerging threats while mitigating catastrophic forgetting and preserving general-purpose model capabilities. Across lifelong prompt injection attack streams, \textsc{COPA} reduces attack success rate by up to $6.3\times$ and $4.4\times$ on average compared to state-of-the-art defenses. These results highlight continual preference optimization as an effective paradigm for defending LLMs against adaptive adversaries.
\end{abstract}



\section{Introduction}
As large language models (LLMs) are increasingly deployed in enterprise settings, they are being integrated into complex tool-augmented workflows that significantly expand their security attack surface~\cite{ravindran2025utdmf, ferrag2024generative}. Despite their widespread adoption in production systems, securing these deployments remains an open challenge. Recent studies demonstrate that widely used systems such as GitHub Copilot and Microsoft Copilot remain vulnerable to carefully crafted prompt injection attacks that can manipulate downstream model behavior~\cite{skillinjection2025, echolean2025}. Reflecting this risk, OWASP consistently ranks prompt injection as the top security vulnerability in LLM-based applications across recent editions of its threat taxonomy~\cite{owasp_llm_top10}.

\begin{figure}[]
    \centering
    \includegraphics{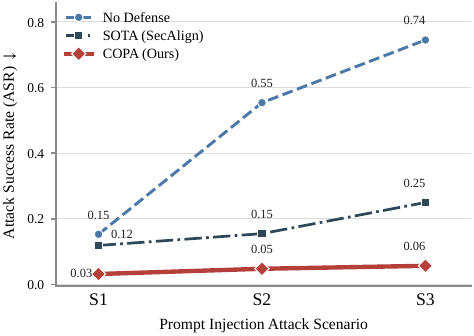}
    \caption{Attack success rate (ASR; lower is better) under three
    escalating prompt injection scenarios (S1--S3). \textsc{COPA} holds ASR
    below 0.06 throughout, while the undefended model and prior defense
    degrade to 0.74 and 0.25 respectively at S3.}
    \label{fig:asr_lifelong}
\end{figure}

Existing defenses against prompt injection attacks typically rely on
alignment~\cite{chen2025struq,chen2025secalign} or classifier-based
filtering~\cite{liu2025datasentinel,inan2023llama}, treating security as a one-time optimization problem. While effective against known attacks, existing defenses struggle to generalize to emerging prompt injection strategies, leaving them vulnerable to adaptive adversaries. As Figure~\ref{fig:asr_lifelong} shows, an undefended Meta
Llama~3.1~8B reaches an ASR of 0.74 on the most challenging scenario, and
the state-of-the-art defense SecAlign~\cite{chen2025secalign} still degrades to 0.25, offering no
consistent protection as attack complexity grows. While recent lifelong safety
alignment frameworks~\cite{wang2026lifelong,li2026lifealign} address evolving
preferences and safety objectives, they are not designed to counter adaptive
prompt injection attacks. This leaves a critical gap: no existing defense
sustains robustness as the attack distribution evolves, motivating a
dedicated continual defense framework.

We address this gap by framing prompt injection defense as \emph{continual security alignment}, a lifelong learning problem in which the adversarial distribution evolves over time and robustness must be preserved across the full history of observed attacks. Unlike prior lifelong alignment settings that focus on shifting user preferences, our setting is inherently adversarial: each new task corresponds to an attack strategy designed to bypass existing safeguards. This distinction has an important consequence: a model must simultaneously retain defenses against previously seen attacks while adapting to new ones, without degrading performance on benign requests. Without jointly satisfying these objectives, continual updates can silently erode earlier defenses even as performance improves on the latest attack distribution, introducing a security-critical form of forgetting that static defenses cannot address.

We propose Continual Preference Optimization for Adaptive Prompt-Injection
Defense (\textsc{COPA}), a lifelong learning framework for prompt injection
defense that adapts sequentially to evolving attack variants. Unlike existing
defenses that optimize against a fixed threat model, \textsc{COPA} formulates
defense as a continual preference optimization problem, implemented as a
lightweight LoRA adapter trained via Group Relative Policy Optimization
(GRPO)~\cite{shao2024deepseekmath} across a sequential curriculum of attack variants, with a
margin-weighted replay buffer that prioritizes examples where the model's
preference margin is low. Our key contributions are threefold:
\textbf{(1) Lifelong defense framework.} We introduce the first continual
learning pipeline for prompt injection defense, enabling sequential
adaptation to new attack types without retraining from scratch while
preserving defenses against previously observed attacks.
\textbf{(2) Margin-weighted experience replay.} We propose a replay buffer
that ranks rehearsal examples by the model's GRPO preference margin,
focusing updates on the attacks the model handles least reliably and
reducing catastrophic forgetting of previously learned defenses. \textbf{(3) Online policy optimization for sequential defense.} We utilize GRPO based online policy updates to adapt the defender at each stage of the attack stream, allowing the model to learn from newly seen attack variants. These contributions shift prompt-injection defense from a static alignment problem to a continual optimization framework that balances robustness, general utility, and preservation. \textsc{COPA} achieves attack success rates up to 6.3$\times$ lower than
state-of-the-art defenses~\cite{chen2025secalign,liu2025datasentinel}, while improving backward transfer by 0.119 and
average performance by 0.156 over the strongest defense baseline, delivering a
defense that strengthens as attacks evolve.

\section{Related Work}
\textbf{Prompt Injection Attacks.}
Prompt injection is widely recognized as a critical vulnerability in LLM-based systems, ranked first in the OWASP Top Ten for LLMs~\cite{owasp2025llmtop10}. These attacks exploit instruction-following behavior to redirect model outputs toward attacker-controlled objectives, either through direct user inputs or indirect injections via external sources such as retrieved documents or tool outputs. Recent studies show that deployed systems remain vulnerable to carefully crafted injection strategies~\cite{skillinjection2025,echolean2025}, while adversarial training on known patterns offers limited generalization to unseen variants~\cite{liu2025secinfer}. Thus, prompt injection is increasingly understood as a dynamic, evolving threat rather than a static failure mode.

\textbf{Defenses Against Prompt Injection.}
Existing defenses against prompt injection attacks fall into two broad categories:
alignment-based approaches, which fine-tune models to distinguish legitimate
instructions from injected ones~\cite{piet2024jatmo,gungor2025eager,chen2025struq, chen2025secalign}, and filtering mechanisms, which detect and block malicious
inputs~\cite{liu2025datasentinel, li2025piguard,hung2025attention,inan2023llama}. While effective against known attack distributions, both categories treat security as a one-time optimization problem and fail to adapt to an evolving threat landscape. Consequently, they remain vulnerable to adaptive adversaries that exploit this static design.

\textbf{Lifelong Safety Alignment.}
Lifelong alignment studies how models can continuously adapt to changing objectives while retaining previously acquired behaviors. Wang et al.~\cite{wang2026lifelong} propose a lifelong safety alignment framework that incrementally updates models to address emerging jailbreak strategies, while LifeAlign~\cite{li2026lifealign} mitigates catastrophic forgetting under sequential alignment tasks~\cite{li2026lifealign}. Dabas et al.~\cite{dabas2025adversarial} further observe that many novel attacks arise from recombinations of previously observed adversarial patterns, suggesting recurring structure within the threat space. However, existing lifelong alignment frameworks assume sequential changes in alignment objectives, whereas prompt injection defense requires adapting to an evolving adversarial distribution. This distinction motivates continual security alignment, where robustness must be maintained against both past and emerging attack strategies.


\textbf{Replay-Based Continual Learning.}
Replay-based strategies are widely used in continual learning to mitigate catastrophic forgetting by storing and revisiting past examples~\cite{li2026lifealign,wang2026lifelong}. In alignment settings, replayed samples typically consist of previously observed preference pairs that help preserve earlier objectives. However, existing methods generally replay samples uniformly, overlooking differences in difficulty and informativeness. This limitation is particularly pronounced under evolving adversarial distributions, where model uncertainty varies substantially across attack types. 

\begin{figure*}[]
    \centering
    \includegraphics[width=.95\textwidth]{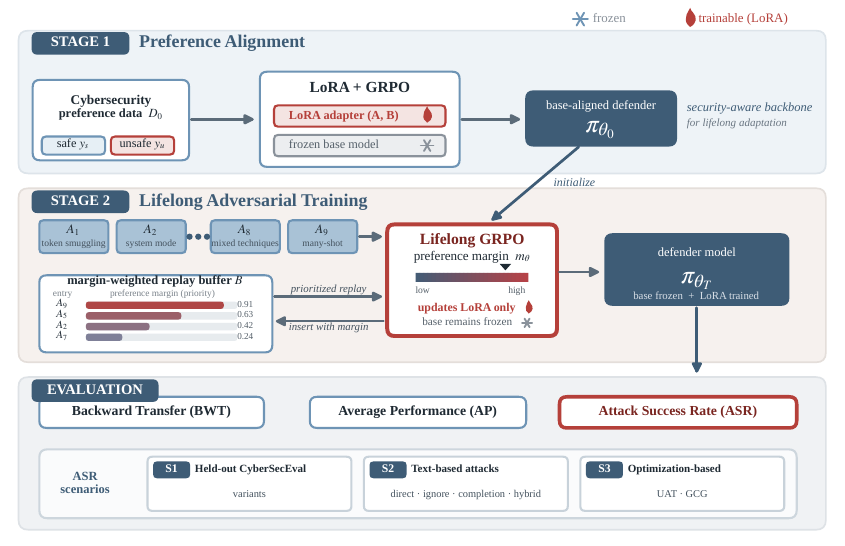} 
    \caption{\textbf{Overview of \textsc{COPA}.} \Design{} aligns a frozen
    base model on safety preferences and then continually adapts a lightweight
    adapter against an evolving curriculum of prompt injection attacks,
    improving robustness over time without forgetting.}
    \label{fig:framework}
\end{figure*}

\section{\Design{} Framework}
Figure~\ref{fig:framework} presents \Design{}, our continual preference optimization framework for defending LLMs against evolving prompt injection attacks. Rather than assuming a fixed attack distribution, \Design{} frames defense as a sequential learning problem in which the defender must adapt to new injection strategies without forgetting how to handle earlier ones. 

The framework comprises four components: (1)~a continual formulation that organizes prompt injection variants into a sequential task stream; (2)~a preference alignment stage that produces a security-aware backbone via GRPO; (3)~a margin-weighted replay buffer that prioritizes preference pairs the defender does not yet handle correctly; and (4)~a lifelong GRPO update that interleaves new-task learning with prioritized replay to mitigate catastrophic forgetting.

\subsection{Problem Formulation}
Let $\mathcal{A}=\{A_1,A_2,\dots,A_T\}$ denote a sequence of prompt injection variants (e.g., \emph{ignore previous instructions}, \emph{token smuggling}, \emph{system mode}). For each variant $A_t$, the defender receives a set of preference pairs
\begin{equation}
\mathcal{D}_t=\{(x^{(t)}_i,\,y^{(t)}_{s,i},\,y^{(t)}_{u,i})\}_{i=1}^{N_t},
\end{equation}
where $x^{(t)}_i$ is an adversarial input drawn from $A_t$, $y^{(t)}_{s,i}$ is a safe response, and $y^{(t)}_{u,i}$ is an unsafe response induced by the attack. The defender $\pi_\theta$ is updated over $T$ stages; after stage $t$ it must defend against $A_t$ as well as all previously observed variants $A_{1:t-1}$. Formally, \Design{} must satisfy three objectives simultaneously:
\begin{enumerate}
\item \textbf{Robustness.} After training on $A_t$, the defender prefers the safe response to the unsafe one on inputs drawn from $A_t$.
\item \textbf{Retention.} After training on $A_{1:T}$, performance on earlier variants $A_1,\dots,A_{T-1}$ does not degrade through catastrophic forgetting.
\item \textbf{Generalization.} The defender resists held-out variants $A_{T+1:T+M}$ that are never seen during training.
\end{enumerate}

\subsection{Threat Model}
We consider an LLM-based defender deployed in a security-sensitive, instruction-following setting where user inputs may contain malicious instructions, placing both the integrity of the model's responses and the reliability of its safety behavior at risk. A successful attack may expose confidential information, elicit unsafe outputs, or override system instructions. 

The adversary is external, with access only to the model's input interface: knowing the prompt format, it injects malicious instructions into the data portion of a query and uses a surrogate LLM to mass-produce adversarial inputs. The adversary cannot modify the defender's weights, replay buffer, training data, or optimization process. Crucially, we assume the attack distribution is \emph{non-stationary}: novel variants $A_{T+1},A_{T+2},\dots$ emerge after deployment, reflecting the real-world evolution of prompt injection strategies~\cite{wang2026lifelong}. \Design{} addresses this setting through continual preference optimization with margin-weighted replay, adapting to new attacks while limiting forgetting of earlier defenses.

\subsection{Stage 1: Preference Alignment}
Before continual training begins, the defender requires a consistent cybersecurity foundation. Different pretrained backbones exhibit varying levels of safety tuning, instruction-following capability, and sensitivity to adversarial prompts. If Stage 2 were applied directly to these heterogeneous models, the lifelong learning signal would be confounded by these initialization differences rather than reflecting true continual adaptation.

Stage 1 mitigates this issue by aligning all backbones to a shared cybersecurity baseline prior to any attack-specific training. We construct a general cybersecurity preference dataset $\mathcal{D}_0$ from the DeepTeam red-teaming framework~\cite{deepteam2025}, consisting of paired benign and adversarial cybersecurity prompts spanning scenarios such as phishing, malware distribution, unauthorized access, and social engineering. Each pair is assigned a binary preference label using an automatic safety judge that determines whether the model response satisfies cybersecurity safety constraints.

We then use $\mathcal{D}_0$ to perform preference optimization via Group Relative Policy Optimization (GRPO)~\cite{shao2024deepseekmath}. We adopt GRPO because group-based normalization stabilizes learning under sparse binary safety rewards and reduces the high-variance gradients of scalar reward estimation, making it well-suited for safety-critical preference optimization with discrete, imbalanced signals. The base model is frozen and a low-rank adapter $\Delta W = AB^\top$, with $A,B\in\mathbb{R}^{d\times r}$ and $r\ll d$, is inserted into each transformer layer; only $(A,B)$ are trained. For each input $x$, GRPO samples a group of $G$ candidate responses $\{o_1,\dots,o_G\}\sim\pi_{\theta_{\text{old}}}(\cdot\mid x)$ and scores each with the binary safety judge, $r_i=R(x,o_i)\in\{0,1\}$ (safe responses receive $1$, unsafe responses $0$). Within each group, rewards are normalized into advantages:

\begin{equation}
\begin{aligned}
\hat{A}_i &= \frac{r_i - \mu_x}{\sigma_x + \epsilon}, \\
\mu_x &= \frac{1}{G}\sum_{j=1}^{G} r_j, \\
\sigma_x &= \mathrm{std}(\{r_j\}_{j=1}^{G})
\end{aligned}
\end{equation}
where $\epsilon > 0$ is a small numerical stability constant preventing division by zero when a sampled group yields uniform safety rewards. The adapter is optimized with the clipped GRPO objective and a KL penalty to a frozen reference policy $\pi_{\text{ref}}$:
\begin{equation}
\begin{aligned}
\mathcal{L}_{\text{GRPO}} &=
- \mathbb{E}_{x \sim \mathcal{X},\,\{o_i\} \sim \pi_{\text{old}}}
\Bigg[
\frac{1}{G}\sum_{i=1}^{G}
\min(\rho_i\hat{A}_i,
\mathrm{clip}(\rho_i)\hat{A}_i) \\
&\quad - \beta D_{\mathrm{KL}}(\pi_\theta\|\pi_{\text{ref}})
\Bigg]
\end{aligned}
\label{eq:grpo}
\end{equation}
where $\rho_i=\pi_\theta(o_i\mid x)/\pi_{\theta_{\text{old}}}(o_i\mid x)$ and $\mathcal{X}$ represents the corresponding prompt distribution ($\mathcal{D}_0$ in Stage 1). Optimizing Eq.~\ref{eq:grpo} yields the base-aligned defender $\pi_{\theta_0}$, a security-aware backbone that serves as the starting point and reference policy for the lifelong stage. Because Stage 1 and Stage 2 operate on disjoint data and address complementary objectives (general alignment versus variant-specific defense), the continual learning dynamics observed in Stage 2 reflect adaptation to evolving attacks rather than uneven baseline alignment.

\subsection{Stage 2: Lifelong Adversarial Training}
Stage 2 adapts $\pi_{\theta_0}$ to the attack stream while preserving earlier defenses. It uses a variant-tagged corpus $\mathcal{D}_1$ labeled with $15$ prompt injection variants from the CyberSecEval benchmark~\cite{wan2024cyberseceval}; the first nine variants form the lifelong training stream $\mathcal{A}=\{A_1,\dots,A_9\}$ and the remaining six are held out for evaluation (Section~\ref{sec:setup}). Training proceeds variant by variant with the same GRPO objective as Stage 1, augmented with a margin-weighted replay buffer.

\textbf{Log-likelihood margin as a replay-priority signal.} GRPO drives the policy update, but to decide \emph{which} past examples to rehearse we score each stored preference pair by a lightweight log-likelihood margin measuring how strongly the current defender favors the safe response over the unsafe one:
\begin{equation}
m_\theta(x,y_s,y_u)=\log\pi_\theta(y_s\mid x)-\log\pi_\theta(y_u\mid x).
\label{eq:margin}
\end{equation}
A large positive $m_\theta$ indicates the defender reliably prefers the safe response, effectively neutralizing the attack; a small or negative $m_\theta$ identifies a pair the defender does not yet handle correctly. Unlike judge- or generation-based scoring, $m_\theta$ requires only two log-probability evaluations on stored text, making it inexpensive to maintain across the buffer.

\textbf{Margin-weighted replay buffer.} \Design{} maintains a per-variant replay buffer $\mathcal{B}=\{\mathcal{B}_1,\mathcal{B}_2,\dots\}$, where each $\mathcal{B}_t$ is a bounded reservoir of pairs from variant $A_t$ and each entry $b=(x,y_s,y_u,m_b)$ stores its tracking margin. Because the policy $\pi_\theta$ continuously shifts during lifelong learning, margins calculated at initialization or insertion become stale over time, failing to reflect subsequent catastrophic forgetting. To ensure prioritization remains sensitive to emerging vulnerabilities, we execute a lightweight margin refresh step over all entries in the historical buffer $\mathcal{B}_{1:t-1}$ at the initiation of each new task stage. 

To focus rehearsal on the attacks the defender handles least reliably, we sample a batch of replay entries $b$ from $\mathcal{B}_{1:t-1}$ via a softmax distribution over their negated margins:
\begin{equation}
p(b) = \frac{\exp(-m_b/\eta)}{\sum_{b' \in \mathcal{B}_{1:t-1}} \exp(-m_{b'}/\eta)}
\label{eq:priority}
\end{equation}
Low-margin pairs, those the defense is currently failing to defend against, are thus replayed more frequently. The temperature $\eta$ controls the sharpness of the prioritization on these hard cases. We bucket replay by variant so that replay is balanced across attack types, allowing the buffer to (i)~preserve accumulated robustness, (ii)~regularize the adapter against overfitting to the current variant, and (iii)~focus capacity on attacks the defender remains vulnerable to. 

\textbf{Lifelong GRPO update.} At task $t$, \Design{} forms a mixed set of prompts from the new-task pairs $\mathcal{D}_t$ and a prioritized replay set $\mathcal{R}_t$ drawn from $\mathcal{B}_{1:t-1}$ via Eq.~\ref{eq:priority}, and applies the GRPO objective over this combination:
\begin{equation}
\mathcal{L}_{\Design}(\theta;t)=\mathcal{L}_{\text{GRPO}}\!\big(\theta;\,\mathcal{D}_t\cup\mathcal{R}_t\big).
\label{eq:copa}
\end{equation}
For every prompt, GRPO samples a fresh group of completions, scores them with the binary judge, and updates only the LoRA adapter $(A,B)$; the base model remains frozen, i.e., $\nabla_{W_{\text{base}}}\mathcal{L}_{\Design}=0$ and $\nabla_{A,B}\mathcal{L}_{\Design}\neq 0$. After task $t$ completes, its pairs $\mathcal{D}_t$ are inserted into $\mathcal{B}_t$ with their freshly observed margins (Eq.~\ref{eq:margin}), and the adapter checkpoint is saved for backward-transfer evaluation.

\section{Experimental Analysis}
\subsection{Experimental Setup}
\label{sec:setup}

\textbf{Experimental Configuration.} Unless otherwise stated, all experiments use Meta-Llama-3.1-8B-Instruct as the backbone model. \Design{} is implemented with LoRA ($r=64$, dropout $=0.1$) and 4-bit NF4 quantization. The initial alignment stage trains on $\mathcal{D}_0$ for one epoch with a learning rate of $2\times10^{-4}$. During continual adaptation, \Design{} performs 30 GRPO updates per task using a learning rate of $1\times10^{-4}$. The replay buffer stores 60 preference pairs. All experiments are conducted on a single NVIDIA A100 GPU.

\textbf{Datasets and Evaluation Protocol.}
\textsc{COPA} is first aligned on $\mathcal{D}_0$, a set of 102
cybersecurity preference pairs from the PromptInjection module of the
DeepTeam Red Teaming Framework~\cite{deepteam2025}. Continual adaptation
then proceeds over a stream of prompt injection variants from the
CyberSecEval benchmark~\cite{wan2024cyberseceval}, ordered by increasing
complexity: nine variants $\mathcal{D}_{1:9}$ form the sequential training
stream, while six disjoint variants $\mathcal{D}_{10:15}$ are held out for
evaluation. All splits are drawn from disjoint prompt pools to prevent
train-test leakage.

We assess robustness under three scenarios of escalating distribution
shift, matching the S1--S3 settings in Figure~\ref{fig:asr_lifelong}.
\textbf{S1 (held-out CyberSecEval variants)} uses the six unseen variants
$\mathcal{D}_{10:15}$ to measure in-family generalization. \textbf{S2 (text-based injections)} evaluates direct, ignore, completion, and hybrid attacks~\cite{liu2024formalizing}, representing common real-world strategies outside the
training family. \textbf{S3 (optimization-based attacks)} evaluates UAT~\cite{liu2024automatic} and
GCG~\cite{li2024faster}, which craft prompts through gradient-based search.

\textbf{Baselines.} We compare \textsc{COPA} against representative state-of-the-art defenses
from two categories. \emph{Detection-based defenses} filter malicious inputs
before they reach the target model: \textbf{LlamaGuard}~\cite{inan2023llama}
uses a separate safety classifier to identify injected instructions, and
\textbf{DataSentinel}~\cite{liu2025datasentinel} fine-tunes a detection LLM
under a minimax objective to flag contaminated inputs. \emph{Preference
optimization defenses} instead align the target model directly:
\textbf{SecAlign}~\cite{chen2025secalign} performs defensive preference
optimization on a static training set.

For a fair comparison, we embed every baseline within the same lifelong
training pipeline as \textsc{COPA}: each method is initialized from the same
backbone, aligned on the same preference dataset $\mathcal{D}_0$, and given
the same opportunity to adapt as new attack variants emerge. The only
difference is the alignment strategy itself, allowing us to isolate the
contribution of our margin-weighted replay and GRPO optimization.

\textbf{Evaluation Metrics.}
We evaluate robustness and knowledge retention with three metrics. For a
test split of variant $A_j$ with $N_j$ adversarial prompts, the attack
success rate (ASR) after training through variant $A_i$ is
\begin{equation}
R_{i,j} = \frac{1}{N_j}\sum_{k=1}^{N_j}\mathbb{1}[\text{success}_k],
\end{equation}
where $\mathbb{1}[\text{success}_k]=1$ when the $k$-th prompt bypasses the
defense. Lower ASR is better: $R_{i,j}=0$ means every attack was defended,
and $R_{i,j}=1$ means every attack succeeded. We report ASR both per variant
and across the sequence.

To quantify retention and overall effectiveness, we adopt the standard
lifelong learning metrics Backward Transfer (BWT) and Average Performance
(AP)~\cite{li2026lifealign}, defined over a performance
matrix $m_{i,j} = 1 - R_{i,j}$, the defense success rate on variant $A_j$
after training through variant $A_i$, where $T$ is the number of variants in
the lifelong sequence.

\textit{Backward Transfer (BWT)} measures whether the defender retains
robustness on earlier variants after the full sequence:
\begin{equation}
\mathrm{BWT} = \frac{1}{T-1}\sum_{j=1}^{T-1}\bigl(m_{T,j} - m_{j,j}\bigr).
\end{equation}
Here $m_{j,j}$ is the defense success rate on variant $A_j$ immediately
after it is learned, and $m_{T,j}$ is the rate on the same variant after the
entire sequence. A positive BWT indicates retained or improved robustness on
earlier variants, while a negative BWT signals catastrophic forgetting. BWT
therefore directly tests whether margin-weighted replay preserves stability.

\textit{Average Performance (AP)} measures cumulative defense quality across
the entire trajectory:
\begin{equation}
\mathrm{AP} = \frac{1}{T}\sum_{k=1}^{T}\frac{1}{k}\sum_{i=1}^{k} m_{k,i}.
\end{equation}
After each stage $k$, performance is averaged over all variants seen so far,
and these per-stage averages are then averaged across the sequence. Higher
AP indicates that the defender sustains strong defense throughout the
sequence, not only at its final state. Unlike a single end-of-training
evaluation, AP penalizes defenders that lose robustness on earlier variants
mid-sequence even if they recover, complementing BWT by capturing the full
trajectory of defense quality.

\textbf{Utility.} To verify that improved prompt injection robustness does not come at the expense of general capability, we evaluate utility
preservation on MMLU-Pro~\cite{wang2024mmlu} and GPQA~\cite{rein2023gpqa}, reporting QA accuracy on both.

\subsection{Results}

\begin{figure}[]
    \centering
    \includegraphics[width=\columnwidth]{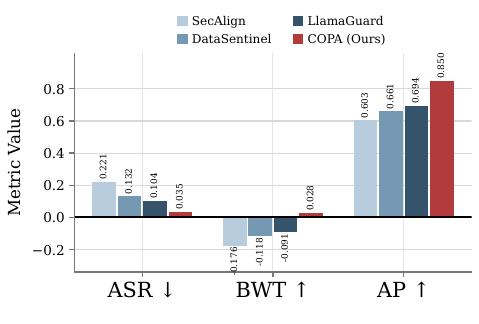}
    \caption{Attack success rate (ASR; $\downarrow$), backward transfer
    (BWT; $\uparrow$), and average performance (AP; $\uparrow$) for \textsc{COPA}
    versus state-of-the-art defenses. \textsc{COPA} is best on every metric
    (ASR 0.035, BWT $+0.028$, AP 0.850), improving robustness significantly.}
    \label{fig:lifelong_results}
\end{figure}

\begin{figure}[]
    \centering
    \includegraphics[width=\columnwidth]{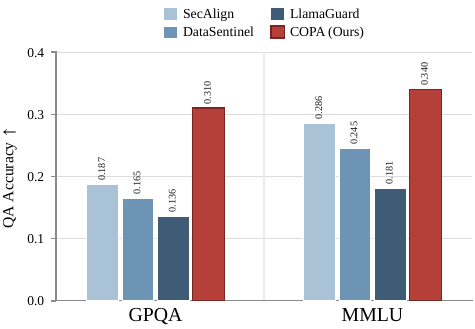}
    \caption{QA accuracy on GPQA and MMLU. \textsc{COPA} achieves the highest accuracy on both, showing that
    \textsc{COPA} improves robustness without degrading general ability.}
    \label{fig:accuracy_results}
\end{figure}

\textbf{Robustness and Knowledge Retention.} Figure~\ref{fig:lifelong_results} compares
\textsc{COPA} against state-of-the-art defenses on attack success rate (ASR),
backward transfer (BWT), and average performance (AP) across the lifelong
attack sequence. \textsc{COPA} achieves the lowest ASR of $0.035$, up to
$6.3\times$ lower than SecAlign (0.221) and $3.0\times$ lower than the strongest baseline (LlamaGuard, 0.104), while DataSentinel (0.132) remains more vulnerable. The root cause of this gap is architectural as static defenses optimize against a fixed snapshot of threat distribution, failing to incorporate new feedback from attack variants after deployment. When the attack distribution shifts at every stage of the lifelong sequence, their learned decision boundaries become misaligned and therefore force ASR to rise. \Design{} avoids this failure mode because GRPO updates continuously reshape the adapter to ensure earlier decision boundaries are reinforced rather than overwritten. \textsc{COPA} also attains the best backward transfer ($+0.028$), indicating that it retains robustness against previously seen attacks rather than overwriting earlier defenses as it adapts, whereas every baseline exhibits negative BWT and thus greater
forgetting. Finally, \textsc{COPA} reaches the highest average performance
($0.850$), confirming that it sustains strong defense throughout the sequence
rather than only at its final state.

\textbf{Utility Comparison.} Figure~\ref{fig:accuracy_results} reports QA
accuracy on GPQA and MMLU for all defenses, measuring whether
improved robustness comes at the cost of general capability. The static
defenses sacrifice substantial utility, with accuracy dropping to as low as
$0.136$ on GPQA and $0.181$ on MMLU. This stems from alignment and
classifier-based defenses shifting the model's output space, suppressing the
very features needed for general reasoning. In contrast, \Design{} attains
the highest accuracy on both benchmarks ($0.310$ on GPQA and $0.340$ on
MMLU), matching or exceeding the undefended base model ($0.300$ and $0.332$).
\Design{} preserves utility by confining updates to a low-rank adapter over a
frozen base and using GRPO's group-relative objective to favor safe
completions over unsafe ones. Unlike prior defenses that buy robustness by
degrading clean performance, \Design{} strengthens defense while preserving
the model's general reasoning ability.

\begin{figure}[]
    \centering
    \includegraphics[width=\columnwidth]{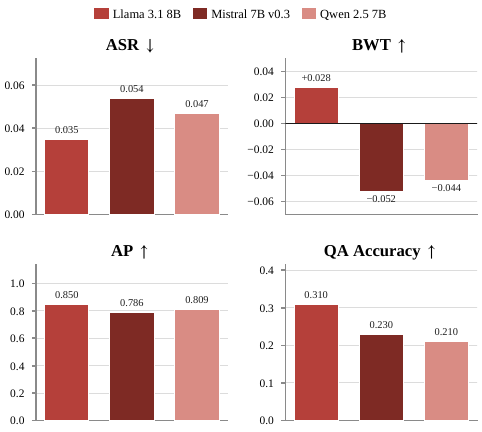}
    \caption{\textsc{COPA} across three LLM backbones (Llama~3.1~8B,
    Mistral~7B~v0.3, Qwen~2.5~7B) on ASR ($\downarrow$), BWT ($\uparrow$),
    AP ($\uparrow$), and GPQA accuracy ($\uparrow$). Consistent performance across
    all backbones indicates that \textsc{COPA} is architecture-agnostic.}
    \label{fig:backbone_results}
\end{figure}

\textbf{Generalization Across LLM Backbones.}
Figure~\ref{fig:backbone_results} evaluates \Design{} on three LLMs (Llama~3.1~8B, Mistral~7B~v0.3, Qwen~2.5~7B) across all four metrics. \Design{} maintains low ASR ($0.035$ to $0.054$), near-zero BWT ($-0.052$ to $0.028$), strong AP ($0.786$ to $0.850$), and QA accuracy of $0.210$ to $0.316$, regardless of the underlying architecture. The three backbones differ in tokenizer design, pretraining corpus, and safety tuning, yet \Design{} delivers consistent gains on each. This suggests the preference margin \Design{} exploits arises from alignment optimization rather than any specific model family. Operating at the adapter level, \Design{} inherits the representation capacity of its backbone without architecture-specific tuning, supporting its use as a model agnostic framework for lifelong prompt injection defense.

\subsection{Ablation Studies} 
\textbf{Effect of margin-weighted replay.} We isolate the contribution of
margin-weighted replay by replacing it with uniform replay while holding the
rest of the \Design{} pipeline fixed: in the full configuration, replay
samples are drawn with probability proportional to a softmax over negated GRPO preference margins so that
low-margin examples are rehearsed more frequently, whereas uniform replay
samples all buffer entries with equal probability. As shown in
Table~\ref{tab:margin_weighted_vs_uniform}, margin-weighted replay achieves a
substantially better balance between adapting to new attacks and preserving
defenses against earlier ones, reducing attack success rate from $0.163$ to
$0.035$ ($-0.128$) and improving backward transfer from $+0.003$ to $+0.028$.
The gap is largest on average performance ($0.850$ vs.\ $0.600$, $+0.250$).
Uniform replay suffers pronounced mid-sequence defense degradation, whereas
concentrating rehearsal on low-margin examples sustains high defense quality
throughout the lifelong sequence.

\textbf{Effect of Training Objective.}
In Table~\ref{tab:training_objective}, we compare DPO and GRPO under the same
lifelong training sequence as \Design{}. Across the full sequence, DPO yields
an ASR of $0.335$, a backward transfer value of $-0.231$, and an average
performance of $0.626$. DPO's offline preference updates overfit the adapter
to the current scenario and overwrite earlier defenses faster than replay can
recover them. In contrast, GRPO remains stable, achieving an ASR of $0.035$,
a backward transfer of $+0.028$, and an average performance of $0.850$.
GRPO's optimization against the model's current behavior plausibly creates a
self-correcting update signal: when the defender becomes uncertain on
replayed examples, the GRPO margin exposes that weakness and guides the
adapter back toward safer behavior without erasing the gains on the current
task. These results reveal that \Design{}'s gains depend not only on replay,
but on pairing replay with an optimizer that can adapt without repeatedly
forgetting.

\begin{table}[]
\centering
\caption{Ablation of the replay strategy in \textsc{COPA}.}
\label{tab:margin_weighted_vs_uniform}
\small
\setlength{\tabcolsep}{6pt}
\renewcommand{\arraystretch}{1.2}
\begin{tabular}{lccc}
\toprule
\textbf{Metric} & \textbf{Margin-Weighted} & \textbf{Uniform} & \textbf{Improvement} \\
\midrule
ASR ($\downarrow$) & $\mathbf{0.035}$ & $0.163$ & $+0.128$ \\
BWT ($\uparrow$)   & $\mathbf{+0.028}$ & $+0.003$ & $+0.025$ \\
AP ($\uparrow$)    & $\mathbf{0.850}$ & $0.600$ & $+0.250$ \\
\bottomrule
\end{tabular}
\end{table}

\begin{table}[]
\centering
\caption{Training objective ablation: GRPO vs. DPO.}
\label{tab:training_objective}
\small
\setlength{\tabcolsep}{6pt}
\renewcommand{\arraystretch}{1.2}
\begin{tabular}{lccc}
\toprule
\textbf{Metric} & \textbf{GRPO} & \textbf{DPO} & \textbf{Improvement} \\
\midrule
ASR ($\downarrow$) & $\mathbf{0.035}$ & ${0.335}$ & $+0.300$ \\
BWT ($\uparrow$)   & $\mathbf{+0.028}$ & $-0.231$ & $+0.259$ \\
AP ($\uparrow$)    & $\mathbf{0.850}$ & $0.626$ & $+0.224$ \\
\bottomrule
\end{tabular}
\end{table}

\section{Conclusion}
We introduced \Design{}, a continual preference optimization framework for
defending LLM-based systems against prompt injection attacks. Rather than
treating defense as a one-time alignment against a fixed attack distribution,
\Design{} formulates it as a lifelong learning problem: it adapts to a stream
of evolving injection variants while a margin-weighted replay buffer
preserves robustness against previously seen attacks. Using the GRPO
preference margin as a label-free signal of defender uncertainty, \Design{}
concentrates rehearsal on the attacks the model handles least confidently.
Across held-out, text-based, and optimization-based scenarios, \Design{}
reduces attack success rate by up to $6.3\times$ relative to state-of-the-art
defenses while achieving positive backward transfer and preserving general QA
accuracy, delivering a defense that holds up as attacks evolve.

\section*{Acknowledgements}
This work has been funded in part by NSF, with award numbers \#2112665, \#2112167, \#2003279, \#2120019, \#2211386, \#2052809, \#1911095 and in part by PRISM and CoCoSys, centers in JUMP 2.0, an SRC program sponsored by DARPA.

\bibliographystyle{IEEEtran}
\bibliography{sample-base}

\end{document}